\documentstyle[epsf,prl,twocolumn,aps,floats]{revtex} 
\begin{document}

\tightenlines
\textheight = 23.97cm

\topmargin = -2.40cm

\overfullrule 0pt

\twocolumn[\hsize\textwidth\columnwidth\hsize\csname @twocolumnfalse\endcsname

\title{ The Solar Neutrino Problem \\ and \\ Gravitationally 
Induced Long-wavelength Neutrino Oscillation
\vglue -2.0cm \hfill \small \bf IFUSP-DFN/99-034\hglue 0.5cm  hep-ph/9909250
\vglue 1.4cm
}  

\author{
A.\ M.\ Gago$^{1,2}$, 
H.\ Nunokawa$^{3,4}$ and 
R.\ Zukanovich Funchal$^{1}$ }

\address{\sl 
$^1$ Instituto de F\'{\i}sica,  Universidade de S\~ao Paulo 
    C.\ P.\ 66.318, 05389-970 S\~ao Paulo, Brazil\\
$^2$  Secc\'{\i}on F\'{\i}sica, Departamento de Ciencias 
Pontificia Universidad Cat\'olica del Per\'u \\
  Apartado 1761, Lima, Per\'u   \\
$^3$ Instituto de F\' {\i}sica Gleb Wataghin, 
    Universidade Estadual de Campinas -- UNICAMP, 
    13083-970 Campinas, Brazil\\
$^4$ Institute for Nuclear Theory, University of Washington, Box 351550
    Seattle, WA 98195, USA  \\
{\rm (September, 1999)}
\vglue -0.6cm
}    

\maketitle
\vspace{.5cm}

\hfuzz=25pt
\begin{abstract} 
We have reexamined the possibility of explaining the solar neutrino 
problem through long-wavelength neutrino oscillations induced 
by a tiny breakdown of the weak equivalence principle of general 
relativity. 
We found that such gravitationally induced oscillations 
can provide a viable solution to the solar neutrino problem.
\vglue -1.35cm
\end{abstract}
\vskip2pc]

\preprint{
\hfill$\vcenter{
                \hbox{\bf IFUSP-DFN/99-XXX} 
                \hbox{\bf hep-ph/9909250}
             }$ }

\newpage

Nature seems to be most strongly in agreement with neutrino 
oscillations. The compelling evidences coming from solar neutrino 
experiments~\cite{homestake,sage,gallex,sk99}, that span over two decades,
and from atmospheric neutrino experiments~\cite{atmospheric} 
are difficult, if not impossible, to be accommodated without admitting 
neutrino flavor
conversion. Nevertheless the dynamics underlying such conversion 
is yet to be established and in particular does not have to be a priori 
related to the electroweak force. 

The interesting idea that gravitational forces may induce neutrino mixing 
and flavor oscillations if the weak equivalence principle
of general relativity is violated, was proposed by Gasperini~\cite{gasper} 
and independently by Halprin and Leung~\cite{hl} about a decade ago, 
and thereafter, many works have been performed on this
subject~\cite{vep,pantaleone,bkl,kuo,fly,lisi,pkm}. 
In Ref.\ \cite{liv} this was shown to be phenomenologically 
equivalent to velocity oscillations of neutrinos due to a possible 
violation of Lorentz invariance~\cite{cg1}. 
So even a tiny breakdown of the space-time structure of special 
and/or general relativity may lead to flavor oscillations 
even if neutrinos are strictly massless.

Some theoretical insight on the type of gravitational potential that 
could  violate the weak equivalence principle can be found in 
Ref.\ \cite{amn}. 
A discussion on the departure from exact Lorentz invariance in the 
standard model Lagrangian  in a perturbative framework 
is developed in Ref.\ \cite{cg2}.

Several authors have investigated the possibility of solving the solar
neutrino problem (SNP) by such gravitationally induced  neutrino 
oscillations~\cite{pantaleone,bkl,kuo}, generally finding it necessary,
in this context, to invoke the MSW like resonance~\cite{hl} since they   
conclude that it is impossible that 
this type of long-wavelength vacuum
%the gravitationally induced  long-wavelength vacuum
oscillation could explain the specific energy dependence 
of the data~\cite{pantaleone,bkl}.

Recently these neutrino oscillation  mechanisms have been 
investigated ~\cite{fly,lisi,pkm}  in the 
light of the experimental results from Super-Kamiokande (SK) on 
the atmospheric neutrino anomaly, obtaining stringent limits 
for the $\nu_\mu \to \nu_\tau$ channel.

We consider in this letter the possibility of explaining 
the most precise and recent solar neutrino data
coming from gallium, chlorine and water Cherenkov detectors
by means of neutrino mixing due to a ``just-so'' 
violation of the weak equivalence principle (VEP).  
We demonstrate that all the data can be well accounted for 
by the VEP induced long-wavelength neutrino oscillation in
contrast to previous conclusions~\cite{pantaleone,bkl,liv}.

We assume that neutrinos of different species will incur different
time delay due to the  weak, static gravitational field in the
intervening space on their way from the Sun to the Earth. 
Their motion in this gravitational field can be appropriately
described by the parametrized post-Newtonian formalism~\cite{mtw} 
with a different parameter for each neutrino type. 
In this manner neutrinos that are weak interaction eigenstates and
neutrinos that are gravity eigenstates will be related by a unitary
transformation that can be parameterized, assuming only two neutrino
flavors,  by a single parameter, the mixing angle $\theta_G$ which can
lead to flavour oscillation~\cite{gasper}. 

Let us briefly revise the formalism that will be used in this work. 
We will assume oscillations only between 
two species of neutrinos, which are degenerate in mass, 
either between active and active ($\nu_e \leftrightarrow
\nu_\mu,\nu_\tau$) or  active and sterile ($\nu_e \leftrightarrow \nu_s$, $\nu_s$ being
an electroweak singlet) neutrinos. 

The evolution equation for neutrino flavors $\alpha$ and $\beta$ propagating 
through the  gravitational potential $\phi(r)$ in the absence of
matter is~\cite{gasper} : 

\begin{equation} 
\hskip  -0.05cm
i{\displaystyle{d \over \displaystyle{dt}}
\left[ \begin{array}{c} \nu_\alpha \\
\nu_\beta \end{array} \right] 
} 
= E \phi(r) \Delta \gamma  \left[ \begin{array}{cc} \cos 2\theta_G &  \sin 2 \theta_G \\
 \sin 2 \theta_G & -\cos 2 \theta_G \end{array} \right] 
\left[ \begin{array}{c} \nu_\alpha \\
\nu_\beta \end{array} \right], 
\label{oscila} 
\end{equation} 
where $E$ is the neutrino energy;
$\Delta \gamma$ is the quantity which measures the magnitude of VEP,
it is the difference of the gravitational couplings between the two 
neutrinos involved normalized by the sum.

There are many possible sources for $\phi$, but it is generally 
believed that the Super Cluster contribution ($\phi \sim 3 \times 10^{-5}$)
would be the dominant one~\cite{kenyon}. Therefore, it    
seems reasonable to ignore any variation of $\phi$ over the whole
solar system and take it as a constant~\cite{form}. 
In this case Eq.\ (\ref{oscila}) can be analytically solved to give
the survival probability of $ \nu_e$ produced in the Sun after
traveling the distance $L$ to the Earth:

\begin{equation}
 P( \nu_e \rightarrow \nu_e) 
= 1 - \sin^2 2\theta_G \sin^2 \frac{\pi L}{\lambda},
\label{prob}
\end{equation}
where the oscillation wavelength $\lambda$ is given by, 
\begin{equation}
 \lambda 
= \left[\frac{\pi {\text{ km}}}{5.07}\right] \left[\frac{10^{-15}}
{|\phi \Delta \gamma|}\right] \left[\frac{ {\mbox{MeV}}}{E}\right],
\label{wavelength}
\end{equation}
which in contrast to the wavelength for mass  induced neutrino 
oscillations in vacuum, is inversely proportional to the neutrino energy.

In this case the survival probability is a function of two unknowns 
parameters that can be fitted, or constrained, by experimental data: 
$\Delta \gamma$ and $\sin 2\theta_G$.
Since the value of the potential $\phi$ in our solar system is somewhat 
uncertain~\cite{form}, we will adopt the procedure used by other authors 
and work with the product $\phi \Delta \gamma$. 

We will perform a fit of the rates and SK recoil-electron spectrum 
but not take into account the day night effect (or zenith
angle dependence) in  SK. 
This is justified  by the fact that day night variations can not be 
induced by this mechanism, and therefore, are irrelevant in determining 
the allowed parameter region. 
We will comment about the possible seasonal variations at the end. 

We first examine the observed solar neutrino rates in the
VEP framework. In order to do this we have calculated the theoretical  
predictions for gallium, chlorine and Super-Kamiokande water Cherenkov  
solar neutrino experiments, as a function of the two VEP parameters, 
using the solar neutrino fluxes predicted by 
the Standard Solar Model by Bahcall and Pinsonneault (BP98
SSM)~\cite{BP98} taking into account the eccentricity of the Earth
orbit around the Sun. 

\begin{figure}
\centering\leavevmode
\epsfxsize=200pt
\epsfbox{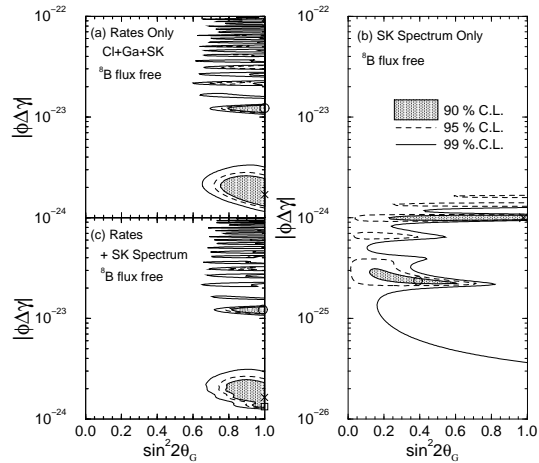}
\vglue -0.01cm
\caption{
Allowed region for $\sin^22\theta_G$ and 
$|\phi \Delta \gamma|$ for (a) the rates only, (b) 
SK spectrum only and (c) rates + SK spectrum 
combined.  
The best fit points are indicated by the crosses
and the local best fit points in the other 90 \% C.L. 
islands are indicated, in each plot, by the open circles. 
The ``test point'' which will be used in Fig. 2 and 4 
is indicated by the open square (see also the text).
}
\label{fig1}
\vglue -0.35cm
\end{figure}

We then have performed a $\chi^2$ analysis to fit these parameters 
and an extra normalization factor $f_B$ for the $^8$B neutrino 
flux, to the most recent experimental results coming from 
Homestake~\cite{homestake} $R_{\text{Cl}}= 2.56 \pm 0.21$ SNU, 
GALLEX\cite{gallex}  and SAGE\cite{sage} combined 
$R_{\text{Ga}}= 72.5 \pm 5.5$ SNU and SK~\cite{sk99} 
$R_{\text{SK}}= 0.475 \pm 0.015$ normalized to BP98 SSM. 
The definition of the $\chi^2$ function to be minimized is 
the same as the one used in Ref.\ \cite{chi2} which essentially 
follows the prescription given in Ref.\ \cite{fogli} 
except that our theoretical estimatives were computed by convoluting 
the survival probability given in Eq.\ (\ref{prob})
with the absorption cross sections taken from Ref.\ \cite{bhp} and 
the  neutrino-electron elastic scattering cross section with radiative
corrections~\cite{xsec} and the solar neutrino flux corresponding to 
each reaction, $pp$, $pep$, $^7$Be, $^8$B, $^{13}$N and  $^{15}$O
and other minor neutrino sources such as $^{17}$F or $hep$ neutrinos
are neglected.   

We will first discuss our results for active to active conversion.
We present in Fig.\ 1 (a) the allowed region determined only by the
rates with free $f_B$ 
and in Table~\ref{tab1} the  best fitted parameters as well 
as the $\chi^2_{\text{min}}$ values for fixed and free $f_B$. 
We found for $f_B=1$ that  
$\chi^2_{\text{min}} = 1.49$ for 3-2=1 degree of freedom and 
for $f_B=0.81$ that $\chi^2_{\text{min}} = 0.32$ for 3-3=0 
degree of freedom. We also have checked that the allowed region 
for fixed $^8$B flux ($f_B=1$) is rather similar to the one 
presented here and so we only give the values of the 
corresponding best fitted parameters for this case 
in Table~\ref{tab1}.

Next we perform a spectral shape analysis fitting the $^8$B 
spectrum measured by SK ~\cite{sk99} using the following 
$\chi^2$ definition:
\begin{figure}
\centering\leavevmode
\epsfxsize=170pt
\epsfbox{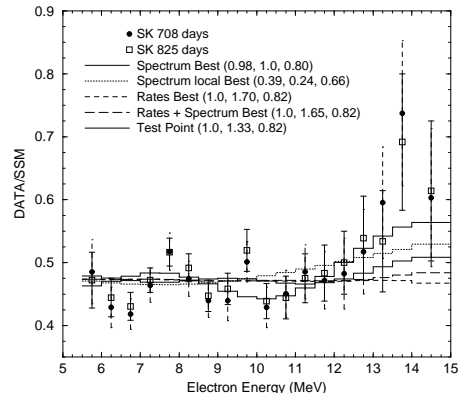}
\vglue -0.01cm
\caption{
Expected recoil-electron spectra at SK for the best fitted parameters 
of the VEP scenarios,  which are indicated in the legend of the 
plot as 
($\sin^2 2\theta_G$,  $|\phi \Delta \gamma| \times 10^{24}$, $f_B$).
The preliminary data from SK are also plotted.}
\label{fig2}
\vglue -0.25cm
\end{figure}
%
%%%%%%%%%%%%%%  Table I  %%%%%%%%%%%%%%%%%%%%%%%%%%%%%%%%%%%%%%%%%%%%%%%%%
%
\begin{table}[h]
\caption[Tab]{The best fitted parameters and $\chi^2_{min}$ for the 
VEP induced long-wave length neutrino oscillation solution to the
SNP. 
The local best fit points in the 2nd 90 \% C.L. islands are indicated 
in the parentheses.
}
\vglue -0.4cm
\begin{center}
\begin{tabular}{ccccc}
Case  & $\sin^2 2\theta_G$ 
& $|\phi \Delta \gamma| \times 10^{24}$ & $f_B$ & $\chi^2_{min}$  \\ \hline
Rates\ ($f_B=1$)    &   1.0\ (1.0)   &1.71\ (12.3)   &  ---&  1.49\ (1.88)  \\ 
Rates  &   1.0\ (1.0)   &1.70\ (12.4)   &  0.81\ (0.81)&  0.32\ (0.71)  \\ 
Spectrum   &   0.98\ (0.39) &1.00\ (0.24)   &  0.80\ (0.66)&  15.8\ (19.8)  \\ 
Combined    &   1.0\ (0.99)   &1.65\ (12.2)   &  0.82\ (0.82)&  22.0\ (23.0)  \\ 
\end{tabular}
\end{center}
\label{tab1}
\vglue -0.8cm
\end{table}
\begin{equation} 
\chi^2 = \sum_i 
\left[\frac{S^{\text{obs}}(E_i)-f_B S^{\text{theo}}(E_i)}{\sigma_i}\right]^2,
\end{equation}
where the sum is performed over all the 18 experimental points
$S^{\text{obs}}(E_i)$ normalized by BP98 SSM prediction for 
the recoil-electron energy $E_i$, $\sigma_i$ is the total
experimental error and $S^{\text{theo}}$ is our theoretical prediction
that was calculated using the  BP98 SSM $^8$B differential flux, 
the $\nu-e$  scattering cross section~\cite{xsec}, the survival 
probability as given by Eq.\ (\ref{prob}) taking into account 
the eccentricity as we did for the rates, the experimental energy 
resolution as in Ref.\ \cite{res} and the detection efficiency as 
a step function with threshold $E_{\text{th}}$ = 5.5 MeV.

After the $\chi^2$ minimization with $f_B=0.80$ we have obtained 
$\chi^2_{\text{min}}=15.8$ for 18-3 =15 degrees of freedom. 
The best fitted parameters that can be found in Table~\ref{tab1} 
permit us to compute the allowed region displayed in Fig.\ 1 (b). 

Finally we have performed a combined fit of the rates and the spectrum 
obtaining the allowed region presented in Fig.\ 1 (c). Again we can 
read from Table~\ref{tab1}  the best fitted parameters. We observe 
that the combined allowed region is essentially the same as the one 
obtained by the rates alone.  In all cases presented in 
Figs.\ 1 (a)-(c) we have two isolated islands of  90\% C.\ L.\ allowed 
regions. See Table~\ref{tab1} for the fitted values corresponding to
the local minimum in these islands.  We note that only the upper 
corner of the Fig.\ 1 (c), for $|\phi \Delta \gamma | > 2 \times 10^{-23}$ 
and maximal mixing in the $\nu_e \to \nu_\mu$ channel can be excluded
by CCFR~\cite{pkm}, and moreover, there are no restrictions in the 
range of parameters we considered in the case of 
$\nu_e \to \nu_\tau$ or $\nu_e \to \nu_s$ oscillations.

In Fig.\ \ref{fig2} we show the expected recoil-electron spectrum in
SK for various fitted parameters of the VEP solution to the SNP.
We see that the data from the  spectrum alone can be quite well 
described by the VEP oscillation mechanism (thick solid line), 
whereas  the prediction for the best fitted parameters from the 
rates alone and from the combined fit give  flatter curves 
(dashed and long-dashed lines). 
Nevertheless parameters for a ``test point'' taken inside 
the 90 \% C.\ L.\ region of Fig.\ 1 (c) can give rise to some 
spectral distortion (thin solid line).

We have performed the same analyses with rates as well as spectrum 
also for the $\nu_e \to \nu_s$ channel. 
Since the allowed regions as well as the fitted recoil-electron spectra 
obtained in this case are rather similar to the ones for active to
active conversion, we do not present them here but only show the best 
fitted parameters and $\chi^2_{\text{min}}$ values  in Table ~\ref{tab2}.  
Although the spectrum alone gives a comparable fit to the active 
to active case, we see that the rates can not be so well explained 
by this type of scenario and consequently the combination gives 
a worse fit. In spite of that this is still much better than the 
mass induced active to sterile vacuum oscillation solution to SNP. 

\begin{figure}
\centering\leavevmode
\epsfxsize=180pt
\epsfbox{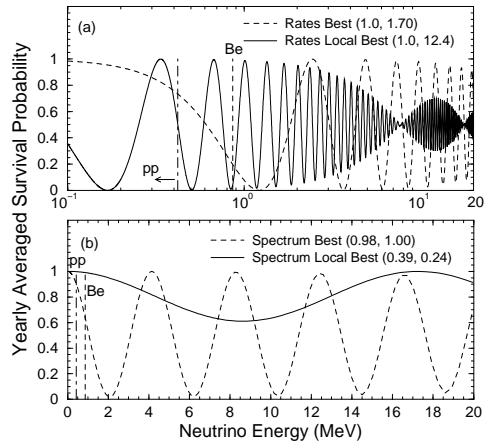}
\vglue -0.01cm
\caption{Yearly averaged survival probability for 
the best fitted parameters, indicated in the parentheses as 
($\sin^2 2\theta_G$,  $|\phi \Delta \gamma| \times 10^{24}$)
in the plots, which can explain well (a) the rates 
or (b) the SK spectrum by the VEP induced neutrino oscillation. 
The energies of the $pp$ as well as $^7$Be neutrinos
are also indicated by the thick dot-dashed and dashed line, 
respectively.} 
\label{fig3}
\vglue -0.4cm
\end{figure}

To understand why it is possible to fit the solar neutrino data we show 
in Fig.\ \ref{fig3} (a) the survival probabilities for the best fitted 
parameters of the VEP induced oscillation.  
Due to the specific energy dependence of the 
probability assumed here  we can actually strongly suppress the Be line 
and still keep the $pp$ neutrino flux high enough to be in agreement
with Ga data, and at the same time obtain $\sim$ 50 \% reduction 
of the $^8$B neutrino flux, which is in fact the required suppression
pattern of the solar neutrino fluxes in order to get a good fit~\cite{mina98}. 
 
Because of the contributions from the strong 
smearing in energy of the scattered electron and of the finite 
experimental energy resolution, the probability alone 
can not give us a precise insight on the spectral 
shape. We can only qualitatively expect 
some distortion for the 
probability in  Fig.\ \ref{fig3} (b).

%%%%%%%%%%%%%%  Table II  %%%%%%%%%%%%%%%%%%%%%%%%%%%%%%%%%%%%%%%%%%%%%%%%%
\begin{table}[h]
\caption[Tab]{Same as Table I but for 
the case of $\nu_e \to \nu_s$ conversion.}
\vglue -0.4cm
\begin{center}
\begin{tabular}{ccccc}
Case  & $\sin^2 2\theta_G$ 
& $|\phi \Delta \gamma| \times 10^{24}$ & $f_B$ & $\chi^2_{min}$  \\ \hline
Rates\ ($f_B=1$)    &   1.0\ (1.0)   &1.80\ (12.1)   &  ---&  3.06\ (3.87)  \\ 
Rates  &   1.0\ (1.0)   &1.80\ (12.1)   &  0.94\ (0.94)&  2.96\ (3.85)  \\ 
Spectrum   &   0.88\ (0.33) &1.01\ (0.24)   &  0.84\ (0.66)&  15.6\ (19.7)  \\ 
Combined    &   1.0\ (1.0)   &1.66\ (12.5)  &  0.94\ (0.94)&  24.7\ (26.2)  \\ 
\end{tabular}
\end{center}
\label{tab2}
\vglue -0.7cm
\end{table}
%%%%%%%%%%%%%%%%%%%%%%%%%%%%%%%%%%%%%%%%%%%%%%%%%%%%%%%%%%%%%%%%%%%

%%%%%%%%%%%%%%%%%%%%%%%%%%%%%%%%%%%%%%%%%%%%%%%%%%%%%%%%%%%%%%%%%%%

\begin{figure}
\vglue -0.3cm
\centering\leavevmode
\epsfxsize=180pt
\epsfbox{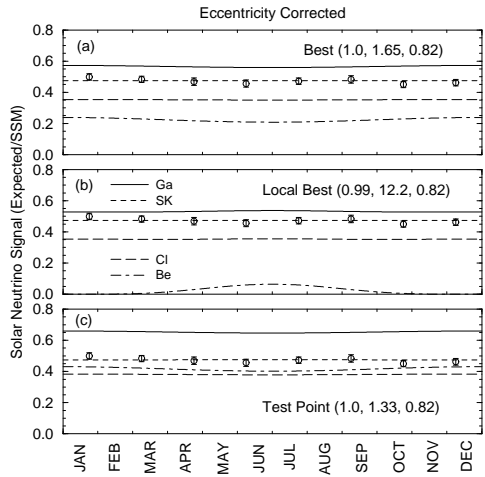}
\vglue 0.2cm
\caption{Expected seasonal variations for the 
fitted parameters of VEP scenarios, 
indicated in the parentheses as 
($\sin^2 2\theta_G$,  $|\phi \Delta \gamma| 
\times 10^{24}$, $f_B$) in each plot.
The preliminary data from SK are also plotted. 
Variations due to the eccentricity of the Earth orbit 
($ \sim 1/L^2$) were subtracted.}
\label{fig4}
\vglue -0.3cm
\end{figure}

Finally, let us discuss about the seasonal variation of the 
solar neutrino signal. 
In contrast to the usual vacuum oscillation
solution to the SNP, in this
scenario, no strong seasonal effect is expected in any of the present 
or future experiments, even the ones that will be sensitive to  
$^7$Be neutrinos such as Borexino~\cite{borexino} and Hellaz~\cite{hellaz}. 
Contrary to the usual vacuum oscillation case, 
the oscillation length for the low energy
$pp$ and $^7$Be neutrinos are very large, comparable to or only a few times
smaller than the
Sun-Earth distance, so that the effect of the eccentricity in the
oscillation probability is small. 
On the other hand, for higher energy neutrinos relevant for 
SK, the effect of the eccentricity in the probability could be large, but 
averaged out after the integration over a certain neutrino energy range. 
These observations are confirmed in Fig.\ \ref{fig4} where we
present the expected seasonal variations for the best fitted
parameters of the VEP induced oscillation scenario. 

%%%%%%%%%%%% Fig. 4 %%%%%%%%%%%%%%%%%%%%%%%%%%%%%%%%%%%%%%%%%%%%%%%%%%%%
%%%%%%%%%%%%%%%%%%%%%%%%%%%%%%%%%%%%%%%%%%%%%%%%%%%%%%%%%%%%%%%%%%%%%

In conclusion we found a new solution to the SNP which is comparable  
in quality of the fit to the other suggested ones.

%%%%%%%%%%%%%% Thanks

We thank Plamen Krastev, Eligio Lisi, George Matsas, Hisakazu
Minakata, Pedro de Holanda and GEFAN for valuable discussions and useful comments.
We also thank Michael Smy for useful correspondence. 
H.N. thanks Wick Haxton and Baha Balantekin and the Institute for
Nuclear Theory  at the University of Washington for their hospitality 
and the Department of Energy for partial support during the 
final stage of this work. 
This work was supported by the Brazilian funding agencies FAPESP and 
CNPq. 
%Funda\c{c}\~ao de Amparo \`a Pesquisa do Estado de S\~ao Paulo (FAPESP) 
%and by Conselho Nacional de Ci\^encia e Tecnologia (CNPq).
%
%%%%%%%%%%%%%% References 

\end{document}